\documentclass[aps,prl,floats,twocolumn]{revtex4}
\usepackage{graphicx}
\begin{document}


\newcommand{\cu} {$^{63}$Cu}
\newcommand{\ox} {$^{17}$O}
\newcommand{\cuo} {CuO$_2$}
\newcommand{\cutp} {Cu$^{2+}$}
\newcommand{\ybco} {YBa$_{2}$Cu$_3$O$_{6+x}$}
\newcommand{\ybcoeight} {YBa$_{2}$Cu$_4$O$_{8}$}
\newcommand{\ybcoca} {Y$_{1-x}$Ca$_x$Ba$_2$Cu$_3$O$_6$}
\newcommand{\etal} {{\it et al.}}
\newcommand{\ie} {{\it i.e.}}
\newcommand{\jpsj} {{J. Phys. Soc. Jpn.}}


\title{Comment on "Localized behavior near the Zn impurity in \ybcoeight~as measured by
nuclear quadrupole resonance"}

\author{M.-H. Julien$^1$, Y. Tokunaga$^{2,*}$, T. Feh\'er$^{2,\ddag}$, M. Horvati\'c$^{2}$, C. Berthier$^{1,2}$}
\affiliation{$^1$Laboratoire de Spectrom\'etrie Physique,
Universit\'e J. Fourier, BP 87, 38402, Saint Martin d'H\'eres,
France} \affiliation{$^2$Grenoble High Magnetic Field Laboratory,
CNRS and MPI-FKF, BP 166, 38042 Grenoble Cedex 9, France}

\date{\today}

\begin{abstract}

Williams and Kr\"amer [Phys. Rev. B {\bf 64}, 104506 (2001)] have
recently argued against the existence of staggered magnetic
moments residing on several lattice sites around Zn impurities in
YBCO superconductors. This claim, which is in line with an earlier
publication by Williams, Tallon and Dupree~[Phys. Rev. B {\bf 61},
4319 (2000)], is however in contradiction with a large body of
experimental data from different NMR groups. On the contrary, the
authors argue in favor of a very localized spin and charge density
on Cu sites first neighbors to Zn. We show that the conclusions of
Williams and Kr\"amer arise from erroneous interpretations of NMR
and NQR data.

\end{abstract}
\maketitle

\section{introduction}

In a recent paper \cite{Williams01}, Williams and Kr\"amer
(hereafter WK) report on the nuclear quadrupole resonance (NQR)
study of a new Cu line in the Zn-doped high-$T_c$ superconductor
(HTSC) \ybcoeight. This line was discovered by Williams
\etal~\cite{Williams00} and was confirmed by Itoh
\etal~\cite{Itoh01,Itoh03}, while it is also possibly visible in
an earlier report by Yamagata \etal~\cite{Yamagata91}. WK argue
that this resonance arises from the four Cu first neighbors
(Cu$_{\rm NN}$) of each Zn impurity in \cuo~planes. To our
knowledge, this is the first time that this very special site can
be resolved through a well-defined line in a Cu NQR spectrum. This
isolated Cu resonance might thus be (somehow) equivalent to the
$^{89}$Y NMR satellite discovered by Mahajan \etal~in 1994
\cite{Mahajan94,Williams95}. The work of WK is potentially
important because the immediate vicinity of Zn is crucial for
understanding impurities in cuprates. For example, while the
presence of staggered magnetic moments around impurities has been
widely established
\cite{Walstedt93,Bobroff97,Julien00,Ouazi03,Tokunaga03}, the
moment on Cu$_{\rm NN}$ has sometimes been described as a local
moment giving rise to the surrounding staggered
response~\cite{Mahajan94,Bobroff97}, while others have considered
Cu$_{\rm NN}$ simply as the first site where the staggered
response appears, in reaction to the broken translational
symmetry of AF couplings~\cite{Julien00}. The two descriptions are
probably indistinguishable from an experimental point of view, as
they both support a very large moment on Cu$_{\rm NN}$ and an
extended staggered polarization with some decay as a function of
distance from the impurity. Despite these differences in the
descriptions, it is important to note that there is no
disagreement between various linewidth data sets from the
different NMR groups.

Nevertheless, Williams and Kr\"amer argue in favor of a very
localized spin and charge density on Cu sites which are first neighbors
of Zn atoms, and against the existence of the staggered magnetic
moments~\cite{Williams01}. These statements, following earlier
claims by Williams \etal~\cite{Williams00}, are in contradiction
with the facts established by the rest of the NMR community
\cite{Walstedt93,Bobroff97,Julien00,Ouazi03,Tokunaga03}. We show
here that the conclusions of WK cannot be sustained by their NQR
measurements. Other comments on the work of WK, complementary
those presented here, can be found in the comprehensive study of
Itoh \etal~\cite{Itoh03}.

\section{NQR Spectroscopy}

Williams and Kr\"amer state that the observation of a resolved
line implies that there is "very localized charge and spin on the
Cu sites that are nearest neighbor to the Zn impurity". We
disagree with this view.

First, it is not possible to address the problem of the spin
density here: because NQR is performed in zero external magnetic
field, there is no measurable staggered magnetization thus no
NQR line-broadening (unless magnetic moments are partially frozen
on the time scale of the experiment, which is not the case in the
temperature range investigated in Ref.~\cite{Williams01}).

Second, the conclusion of WK regarding the charge density is based
on the assumption that "for small changes the difference between
the \cu~NQR frequency at the Cu$_{\rm NN}$ and Cu$_{\rm NNN}$
sites is proportional to the change in the hole concentration".
This assumption is quite questionable. While the \cu~NQR frequency
$^{63}\nu_{\rm NQR}$ can indeed be related to the average on-site
hole density \cite{Zheng95,Gippius97}, this does not mean that the
"lattice" contribution (charge from surrounding sites
\cite{Shimizu93}) is negligible. As an experimental
counterexample, Zn-doping in undoped La$_2$CuO$_4$ induces
modifications of the Cu NQR frequency \cite{Carretta97}, which are
comparable to the 4\% relative change observed here by Williams
and Kr\"amer \cite{Williams01}. Because there are no doped holes in
La$_2$CuO$_4$, the effect must come from a "lattice" contribution.
The substitution of a Cu atom by Zn is expected to modify the
electric field gradient tensor at the Cu$_{\rm NN}$ site, as both
the lattice contribution and the local symmetry change. Hence,
there is no {\it a priori} reason for the Cu$_{\rm NN}$ site to
have almost the same NQR frequency as the sites far from the
impurity. For example, recent state-of-the-art calculations of
electric field gradients by Bersier \etal~show that the change
in$^{63}\nu_{\rm NQR}$ is readily explained by a small shift
of the O$_{\rm NN}$ position~\cite{Bersier03}.

\section{\cu~NQR $T_1$}
In Zn-doped YBCO, the spin-lattice relaxation rate of $^{63}$Cu
nuclei (1/$^{63}T_1$) does not show the characteristic drop
observed below 150~K in pure samples (pseudogap behavior), but
continues to increase with decreasing $T$
\cite{Williams01,Zheng96,Itoh01,Julien00}. WK correctly point out
that this increase does not {\it necessarily} imply an enhancement
of AF correlations. However, inelastic neutron scattering
measurements have demonstrated that the enhancement occurs only
for $q$ close to the AF wave-vector $(\pi/a,\pi/a)$ and only for
energies $\omega$ below the pseudogap energy scale
\cite{neutrons}. This means that Zn-doping does enhance AF
fluctuations at low energy. In this context the expression
"enhanced AF correlations" (introduced for spin chains
\cite{Balster97}) simply means "enhanced staggered magnetization",
which is the zero-frequency limit of the AF spin fluctuations.

According to WK, it is "unlikely that the 1/$^{63}T_1T$ data can
be interpreted within the enhanced antiferromagnetic correlation
model [...]. Rather, 1/$^{63}T_1T$ at low $T$ in substituted
samples just appears to be a continuation of the Curie-like
behavior observed in the pure materials for high temperatures". To
our knowledge, there is no theoretical argument according to which
the dynamics of the staggered moments \cite{Balster97} is
incompatible with a Curie-like behavior for 1/$^{63}T_1T$, should
this behavior already be present at high $T$ or not. Actually, in
the context of the cuprates, the Curie-Weiss behavior is even
suggestive of AF correlations: it is precisely in this way that
the high-$T$ behavior of $^{63}T_1$ has been interpreted by the
entire NMR community to date. A smooth evolution from the pure to
the Zn-doped materials would thus not be surprising, as there is
no difference in the bare electronic {\it structure} between sites
where the magnetization is enhanced and those where it is reduced:
One always deals with the same correlated Cu$^{2+}$ moments and
the bare AF coupling remains the same, only the on-site
magnetization changes smoothly as a function of position in the
plane and as a function of $T$. We also note that for a magnetic
correlation length of 2-3 lattice spacings, a sizeable
magnetization exists on most sites for Zn-doping values of only a
few percent. Thus it is probably inappropriate to think in terms
of separate dynamics for the first neighbors of each impurity and
for the "bulk" (although some spatial inhomogeneity in the spin
dynamics probably exists even within this purely magnetic picture
\cite{remT1}).

\section{\cu~NQR $T_2$}

Williams and Kr\"amer find that, for Cu sites which are {\it not}
nearest neighbors to Zn, $^{63}T_{2G}$ is close to the value
obtained in the pure compound. They infer that the spins remain
"like", and they conclude that "this provides further evidence,
within the MMP model, that there is no suppression or enhancement
of AF correlations for distances greater than one lattice
parameter away from the Zn impurity".

We disagree with this, for several reasons. First, it must be
emphasized that it has always been believed in cuprates that
"like" spins do not contribute to spin-spin relaxation, and
formulas that are used for the interpretation of $T_{2G}$ take
into account only the $I^i_zI^j_z$ terms and not the
I$^i_+$I$^j_-$ ones \cite{Pennington91}. Therefore, any
distribution of local magnetization should not affect $T_{2G}$
directly. Second, the NQR line, at variance with the NMR one, is
not broadened by the staggered magnetization around the Zn ions,
because in zero external field no static local field (on the NQR
time scale) is induced. Thus the arguments of Williams and Kr\"amer
would not apply to their own data. Finally, as already stated
above, the expression "enhanced AF correlations" has been used for
an enhancement of staggered magnetization around a Zn impurity,
which is related directly to the real part of the spin
susceptibility $\chi(q)$ at $q=Q_{AF}$ \cite{Julien00}. The
asymptotic part of $\chi^\prime(Q_{AF})$, which is responsible for
the Cu, O and Y NMR broadening, has always been estimated with the
value of $\chi(q)$ {\it in the host}
\cite{Walstedt93,Bobroff97,Julien00}. This same, unchanged value
also determines $T_{2G}$.

\section{Compatibility between $^{89}$Y, \cu~and $^{17}$O NMR
data}

\begin{figure}[t!]
\centerline{\includegraphics[width=3.2in]{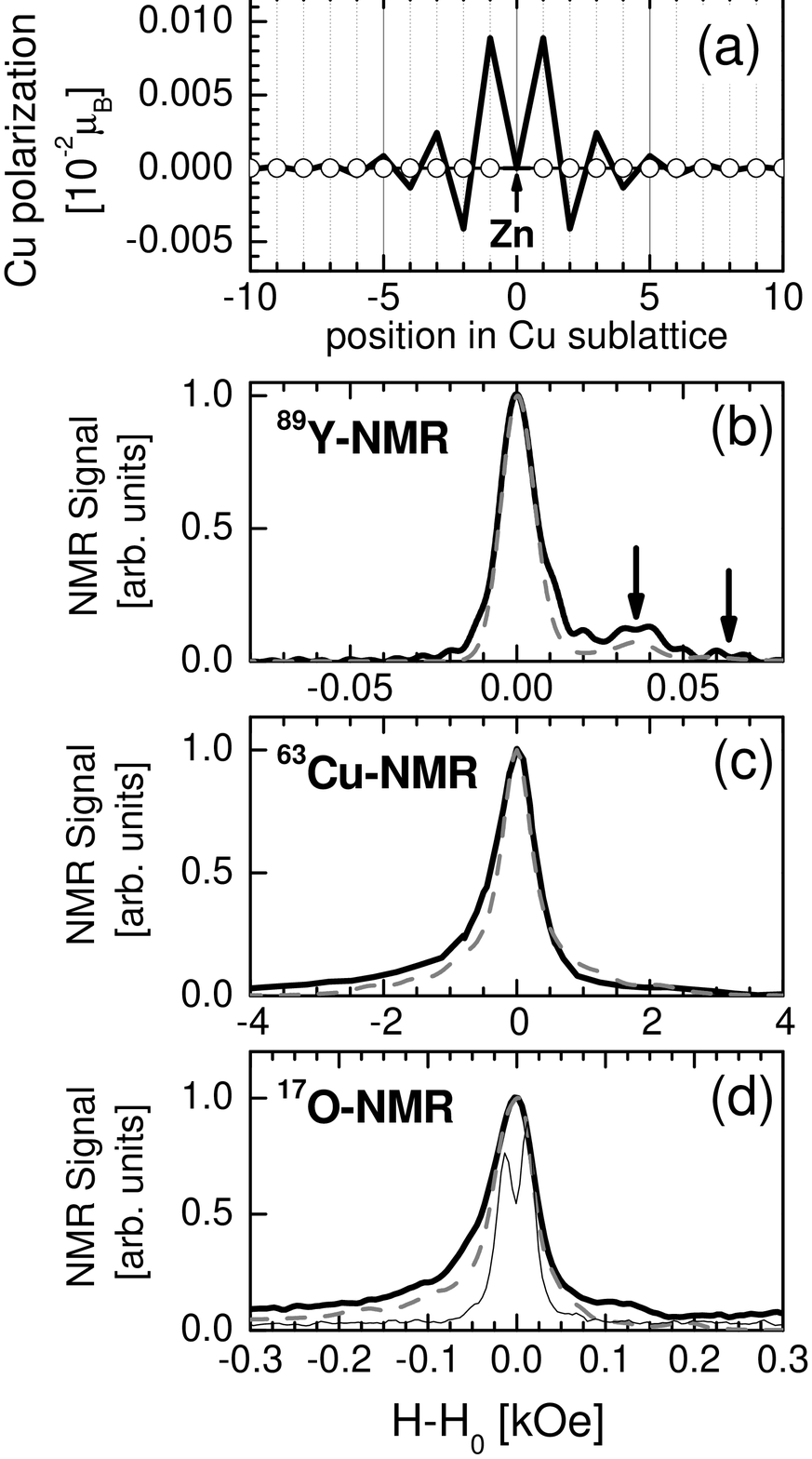}} \caption{(a)
Quantitative model of staggered magnetization around a Zn impurity
(1D cut for clarity; details in~\cite{Tokunaga03}). (b), (c) and
(d) $^{89}$Y, \cu, and \ox~NMR lines in \ybcoeight, doped with 1\%
of Zn per planar Cu, at $T$=50~K and with $H_0(||c)$=14.0~T
(continuous lines, from \cite{Tokunaga03}), together with the
hyperfine field distribution (dashed lines) computed from the
model in (a) using hyperfine coupling constants taken from the
literature.}
\end{figure}

The last argument against an enhanced staggered magnetization,
invoked in \cite{Williams01} but previously developed in
\cite{Williams00}, is that $^{89}$Y NMR spectra show two
Zn-induced lines, both at a frequency lower than that of the main
line, while Williams and coworkers expect additional lines on both
sides of the main line or a symmetric broadening of this main
line. Williams and Kr\"amer conclude that "It is not possible to
account for the $^{89}$Y NMR data within the enhanced
antiferromagnetic correlations model".

In this paragraph we show, only from a qualitative inspection of
NMR data, that this statement is not correct. The location of
$^{89}$Y satellite lines indicates that the magnetic moment on
Cu$_{\rm NN}$ and on Cu$_{\rm NNN}$~(the next nearest neighbors)
is much larger than that on sites located further from the
impurity. However, the main $^{89}$Y NMR line (sites which are
neither Cu$_{\rm NN}$ nor Cu$_{\rm NNN}$) definitely broadens with
Zn doping \cite{Mahajan94}. This demonstrates immediately that the
perturbation is not limited to Cu$_{\rm NN}$ and Cu$_{\rm NNN}$,
but affects the "bulk" as well, albeit with a lower magnitude. In
the example of \cu~NMR (this nucleus has a much larger hyperfine
coupling and a smaller averaging effect than $^{89}$Y
\cite{couplings}), in underdoped \ybco, the width of the \cu~ NMR
central line increases by a factor of 5 between room temperature
(RT) and 80~K, and by a factor of 10 between RT and 24~K
\cite{Julien00}. It is impossible to explain such a strong effect
with a perturbation which does not extend further than the first
and second neighbors to Zn (these represent less than 15\% of the
total number of sites in Ref.~\cite{Julien00} with 1.5\%
Zn/Cu(2)). One may wonder why the Cu$_{\rm NN}$ and Cu$_{\rm
NNN}$~sites, which cause isolated $^{89}$Y NMR lines, have never
been identified in \cu~and $^{17}$O NMR spectra. In fact, because
of the large moment on Cu$_{\rm NN}$, such lines must be severely
shifted in the tails of broad and complex (quadrupolar split)
spectra, making them difficult to identify. Furthermore, Cu$_{\rm
NN}$ and O$_{\rm NN}$ sites may experience wipeout and changes in
both the quadrupole and hyperfine coupling tensors. Because the
magnitude of these changes remains unknown, it is difficult to
predict accurately where these weak resonances should be located.
Recently, Ouazi \etal~also noted that different nuclei probe the
polarization at different length scales~\cite{Ouazi03}.

Next we show briefly that calculations provide quantitative confirmation
of the staggered magnetization model. Fig.~1 shows \cu, \ox~and $^{89}$Y
NMR data taken in the same \ybcoeight~sample, at the same
temperature ($T$=50~K) and in the same magnetic field
(H$_0$=14.0~T). The magnetization profile of a model of staggered
polarization is shown in the upper panel. It is clear that the
distribution of hyperfine fields computed from this model explains
quantitatively the \cu, \ox, and $^{89}$Y data (\cite{Tokunaga03},
details will be published elsewhere). Remarkably, a staggered
magnetization including a large moment on Cu$_{\rm NN}$, combined
with the hyperfine coupling of $^{89}$Y, produces two satellite
lines, both on the high field (low frequency) side of the main
line (arrows in Fig.~1). These results, obtained in a
stoichiometric compound, are in agreement with the interpretation
of $^{89}$Y NMR spectra accepted for years~\cite{Mahajan94} and with
the recent work of Ouazi \etal~in \ybco~\cite{Ouazi03}.

\section{Other remarks}

The enhancement of the staggered magnetization on many Cu sites
around Zn impurities in YBCO is supported by $^{63}$Cu, $^{89}$Y
NMR and $^{17}$O NMR measurements
\cite{Mahajan94,Walstedt93,Bobroff97}. We are not aware of any
experimental report conflicting with these data. A proof of the
staggered character of the spin polarization (already suggested in
\cite{Walstedt93,Bobroff97}) was proposed in \cite{Julien00}. To
our knowledge, no counterargument has yet been put forward.

Williams and Kr\"amer suggest that the results of
Ref.~\cite{Julien00} might be questionable because the contribution
to the signal from the Cu chains was subtracted. However, both
the raw data (Fig.~1(c) in Ref.~\cite{Julien00}) and the discussion
of Ref.~\cite{Julien00} demonstrate that the Cu(2) line-broadening
is undoubtedly {\it not} related to the Cu(1) chain signal.
Furthermore, Fig.~1 shows results for Zn-doped \ybcoeight, in which
there is no overlap between Cu(2) and Cu(1) signals \cite{Tokunaga03}.

The GHMFL is a "Laboratoire conventionn\'e aux universit\'es
J.~Fourier et INPG Grenoble I". The Laboratoire de Spectrom\'etrie
Physique is supported by grant UMR CNRS N°5588.


\begin{references}
\bibitem{remT1} Additional inhomogeneity and complexity is likely to
arise from modifications to the distribution and to the dynamics of
the doped holes.
\bibitem{couplings} There is substantial averaging effect for $^{89}$Y
NMR because each $^{89}$Y nucleus is coupled to eight Cu sites (four
in each \cuo~plane), while \cu~is coupled to five in the same
plane. For \cu~, the signs of the couplings cause cancelation on
averaging \protect\cite{Julien00} for $H||c$. Planar $^{17}$O is
coupled to two Cu nearest neighbors.
\end{references}
\end{document}